\documentclass[10pt,aps,prd,twocolumn,showpacs,amsmath,nofootinbib]{revtex4-2}

\usepackage[utf8]{inputenc}
\usepackage[american]{babel}
\usepackage[a4paper,margin=1.6cm]{geometry}

\usepackage{amsmath,amssymb}
\usepackage{booktabs}
\usepackage{graphicx}
\usepackage[labelfont=bf,font+=smaller,justification=centerlast]{caption}
\usepackage[labelfont=normal,font=normal,font=small]{subcaption}
\usepackage{slashed}

\usepackage[absolute,overlay]{textpos}

\usepackage{hyperref}
\usepackage{cleveref}

\renewcommand{\cref}[1]{\Cref{#1}}

\Crefname{equation}{Eq.}{Eqs.}
\Crefname{figure}{Fig.}{Figs.}
\Crefname{tabular}{Tab.}{Tabs.}
\Crefname{section}{Sec.}{Secs.}

\hypersetup{%
colorlinks,%
citecolor=blue,urlcolor=blue,linkcolor=blue,%
hypertexnames=false%
}

\newcommand{\p}{p_1}
\newcommand{\pp}{p_2}
\newcommand{\ppp}{p_3}
\newcommand{\pppp}{p_4}

\newcommand{\tz}{(1)}
\newcommand{\ta}{(2a)}
\newcommand{\tb}{(2b)}

\allowdisplaybreaks[0]

\hypersetup{%
pdftitle={Standard Model estimate of K->pi4e branching ratio},%
pdfauthor={T. Husek}%
}

\begin{document}

\title{Standard Model estimate of \texorpdfstring{$K^+\to\pi^+4e$}{K->pi4e} branching ratio}
\author{Tom\'{a}\v{s} Husek}
\email{tomas.husek@thep.lu.se}
\affiliation{Department of Astronomy and Theoretical Physics, Lund University,\\ S\"olvegatan 14A, SE 223-62 Lund, Sweden}
\date{\today}

\begin{abstract}
The branching ratio of the $K^+\to\pi^+e^+e^-e^+e^-$ ($K^+\to\pi^+4e$) decay is calculated at leading order in the Standard Model. The dominance of the neutral-pion pole determines the overall branching ratio to be $B(K^+\to\pi^+4e)=B(K^+\to\pi^+\pi^0)B(\pi^0\to4e)\approx7.0(3)\times10^{-6}$. The significance of this contribution is very much concentrated in the context of the whole available phase space, throughout most of which the one-photon-exchange topology is prevalent in turn. It is thus interesting to present branching ratios for only parts of the allowed kinematical region. We find, for instance, $B(K^+\to\pi^+4e,\,m_{4e}>150\,\text{MeV})=6.0(6)\times10^{-11}$.
\end{abstract}


\maketitle

\begin{textblock*}{\paperwidth}(-\leftmargin,1cm)
   \hfill {\small LU TP 22-46}
\end{textblock*}

\section{Introduction and summary}

The flavor-changing (or, in particular, strangeness-changing) neutral-current weak transitions are absent at tree level in the Standard Model (SM).
At the same time, they are manifest, among others, in radiative non-leptonic kaon decays like $K^+\to\pi^+\ell^+\ell^-(\gamma)$, $\ell=e,\mu$, and have become an interesting probe of SM quantum corrections and beyond.
The related underlying radiative modes (transitions) $K^+\to\pi^+\gamma^*(\gamma)$ have been studied before:
They have been calculated in Chiral Perturbation Theory (ChPT)~\cite{Weinberg:1978kz,Gasser:1983yg,Gasser:1984gg}, enriched with electroweak perturbations~\cite{Ecker:1987qi,Ecker:1987hd}, at leading order (LO) (at one-loop level) and beyond, including the dominant unitarity corrections from $K\to3\pi$~\cite{Gabbiani:1998tj,DAmbrosio:1998gur}.
The long-distance-dominated $K^+\to\pi^+\gamma^*$ transition turns out to be essential also for the decay mode to which this work is dedicated: $K^+\to\pi^+4e$.
Moreover, one needs to consider
the $K^+\to\pi^+\gamma^*\gamma^*$ transition, i.e.\ when both photons are off-shell, which entails a significant challenge.

Out of all possible contributions to $K^+\to\pi^+4e$, the neutral-pion exchange clearly dominates when the $\pi^0$ becomes on-shell.
The overall branching ratio for this 5-track decay is then saturated by the contribution of the associated narrow $\pi^0$ peak and can be directly determined as $\mbox{$B(K^+\to\pi^+4e)$}\allowbreak=\mbox{$B(K^+\to\pi^+\pi^0)B(\pi^0\to4e)$}$.
(This has also been checked here.)
It then becomes challenging to observe the $K^+\to\pi^+4e$ process away from $m_{4e}=M_{\pi^0}$, i.e.\ when the leptons do not originate from the decay of the (nearly) on-shell $\pi^0$.
However, it is exactly the suppressed decay rate that makes such reactions attractive to study, at least from the point of view of beyond-Standard Model (BSM) physics.
It is clear that to identify a new-physics-scenario contribution taking part in a decay, one should have a rough estimate of the SM rate.
Only then the possible new-physics effects can be spotted and become visible as deviations from such SM predictions.

Related to $K\to\pi4e$ decays, however, there has not been any number presented in the literature which would be more than an order-of-magnitude estimate.
In particular, for instance, it is naturally believed that it is unlikely for these branching ratios to exceed the $\mathcal{O}(10^{-10})$ benchmark~\cite{Hostert:2020xku}, since they are suppressed with respect to e.g.\ $B(K^+\to\pi^+e^+e^-)\approx 3\times10^{-7}$ or $B(K^+\to\pi^+\gamma\gamma,\,\text{non-res.})\approx\mbox{$1\times10^{-6}$}$ simply due to phase-space factors and additional QED vertices by $\mathcal{O}(\alpha^2)$.
(Indeed, in the present work, we find that the non-resonant topologies give rise to \mbox{$B(K^+\to\pi^+4e,\,\text{non-res.})=7.2(7)\times10^{-11}$}.)
This leads to increasing attention of theorists, and possible BSM scenarios are being explored.
For example, in Ref.~\cite{Hostert:2020xku}, a model is introduced in which the $K\to\pi4e$ decays proceed via $K\to\pi (X'\to XX)$ intermediate states, with a cascade of dark-sector particles $X^{(\prime)}$ and underlying dynamics potentially significantly enhanced compared to the SM case.
In turn, the searches then follow in suitable experiments, and more precise knowledge of SM background, ideally at the level suited for Monte Carlo (MC) implementation, becomes essential.
This is the main motivation behind the present work.
Besides the expressions for the matrix elements, the squares of which are the vital input for the MC event generators, we also provide here the branching ratios related to the respective contributions for reference.

At present, the $K^+\to\pi^+e^+e^-$ single-event sensitivity at NA62~\cite{NA62:2017rwk} is $\mathcal{O}(10^{-11})$~\cite{NA62:2022tte}.
The acceptance for the \mbox{$K^+\to\pi^+4e$} channel is likely to be lower, leading to $\mathcal{O}(10^{-10})$ or even $\mathcal{O}(10^{-9})$ expected sensitivity, i.e.\ up to two orders of magnitude above the SM estimate presented here.
Thus, unless a possible BSM signal would manifest itself inside the detector, a SM signature of a \mbox{$K^+\to\pi^+4e$} decay outside the $m_{4e}=M_{\pi^0}$ kinematical region should not be observed beyond statistical fluctuations.

Further work is currently in progress: It is planned to add the neutral channel, i.e.\ to treat the branching ratios for \mbox{$K^0\to\pi^04e$}, and to include the expressions for matrix elements squared.

In \cref{sec:theory}, we first present the contributions that appear at LO in the Standard Model and then provide a brief description of how the relevant matrix elements are obtained.
Some suitable ChPT Lagrangians employed to calculate the pion-pole contribution are included.
Section~\ref{sec:results} then combines and compares all the terms, and provides numerical results for respective contributions to the branching ratio, obtained in terms of a simple Monte Carlo method.
Some of the details are deferred to \cref{app:PS} so as not to interrupt the flow of the section.
The one-fold differential decay widths as functions of the 4-lepton invariant mass $m_{4e}$, as well as the branching ratios, are presented in \cref{fig:3} and \cref{tab:1}, respectively.

\section{Theoretical setting}
\label{sec:theory}

\begin{figure}[t]
    \centering
    \hypertarget{1}{}
    \includegraphics[width=0.6\columnwidth]{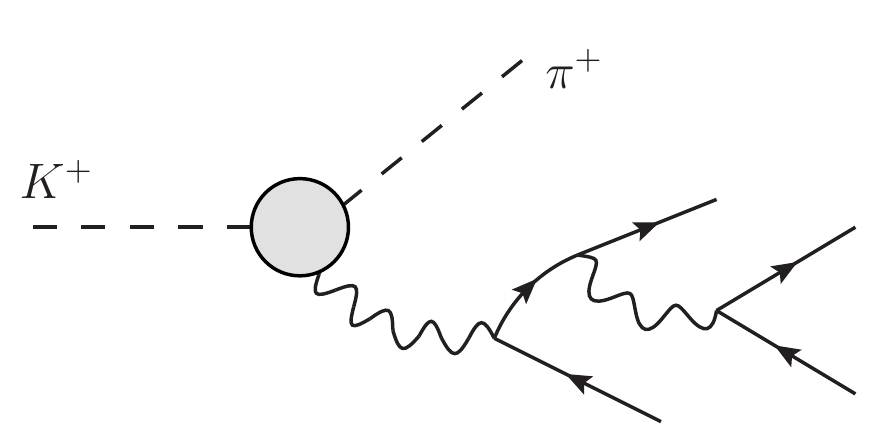}
    \caption{The one-photon-exchange topology.
    There is a cross diagram where the additional off-shell photon is radiated from the positron line (obtained for instance by changing the direction of the fermion-number flow).
    These two then serve as a gauge-invariant building block that comes in 4 permutations of external legs.}
    \label{fig:1}
\end{figure}

\begin{figure}[t]
    \hypertarget{2a}{}
    \hypertarget{2b}{}
    \centering
    \begin{subfigure}[t]{0.45\columnwidth}
    \includegraphics[width=\columnwidth]{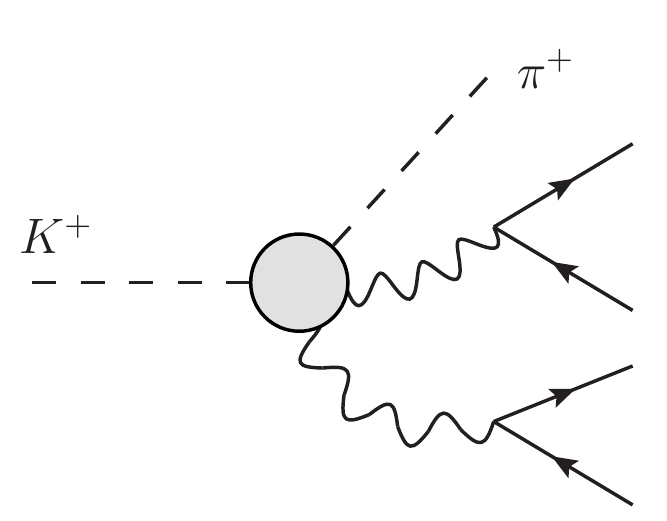}
    \caption{}
    \label{fig:2a}
    \end{subfigure}
    \begin{subfigure}[t]{0.5\columnwidth}
    \includegraphics[width=0.93\columnwidth]{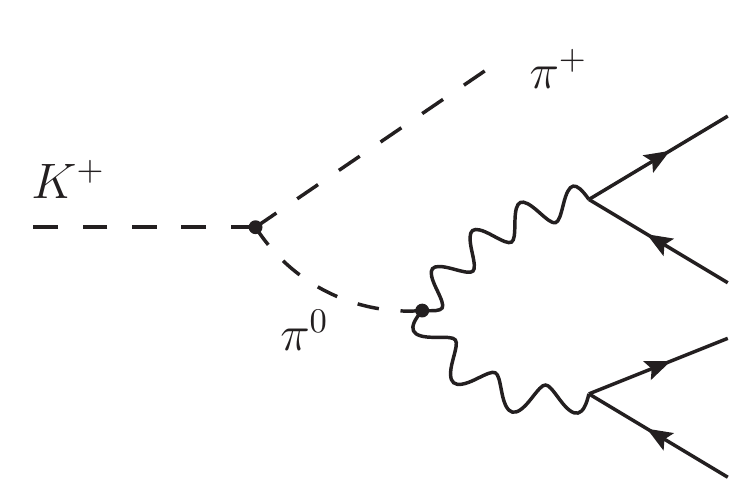}
    \caption{}
    \label{fig:2b}
    \end{subfigure}
    \caption{The two-photon-exchange topology.
    Each of the diagrams comes with one extra permutation of the momenta of the same-charged leptons.}
    \label{fig:2}
\end{figure}

At lowest order in the QED expansion (at $\mathcal{O}(\alpha^2)$) and in ChPT (at $\mathcal{O}(G_\text{F})$, or, equivalently, $\mathcal{O}(G_8)$ and $\mathcal{O}(G_{27})$), there are 3 distinctive underlying topologies present on the meson side of the amplitude:
\begin{enumerate}
    \item $K^+\to\pi^+\gamma^*$ conversion, as shown in \cref{fig:1}, which we will call the single-photon exchange (between the meson and lepton parts);
    \item $K^+\to\pi^+\gamma^*\gamma^*$ conversion, as shown in \cref{fig:2}, which we dub the two-photon exchange, manifesting itself as two sub-topologies:
    \begin{enumerate}
        \item $K^+\to\pi^+\gamma^*$ conversion with an extra radiative photon coming either from the meson legs or from the vertex itself, as shown in \cref{fig:2a},
        \item $K^+\to\pi^+\pi^{0*}$, $\pi^{0*}\to\gamma^*\gamma^*$, as shown in \cref{fig:2b}.
    \end{enumerate}
\end{enumerate}
A few remarks are in place here.
Regarding the topology {\ta}, when one of the two photons becomes on-shell, such a contribution is an important part of the radiative corrections in $K^+\to\pi^+\ell^+\ell^-$ decays~\cite{Kubis:2010mp,Husek:inprep} and is naturally considered when the $K^+\to\pi^+\gamma^*$ transition form factor is measured.
The topology {\tb} might become important for the $K^+\to\pi^+e^+e^-(\gamma)$ decay and the related form-factor extraction, and as a background for the measurement of the branching ratio of the rare decay $\pi^0\to e^+e^-$.

The most nontrivial ingredient of topology {\tz}, the $K^+\to\pi^+\gamma^*$ conversion, has already been extensively studied in the literature, as mentioned in the introduction.
Its contribution can be represented via a single form factor, for which several parametrizations are being used.
In what follows, we denote this form factor simply as $F(s)$, with $s$ being the virtuality of the photon.
It is related to the standard $W_+(z)$, introduced in Ref.~\cite{DAmbrosio:1998gur}, through
\begin{align}
    F(s)
    &=\frac2{(4\pi)^2M_K^2}\,W_+\big(s/M_K^2\big)\,,\\
    W_+(z)
    &=G_\text{F}M_K^2(a_++b_+z)+W_+^{\pi\pi}(z)\,.\label{eq:W+}
\end{align}
For the parameters in \cref{eq:W+}, we employ $a_+=-0.584(8)$ and $b_+=-0.700(35)$, the weighted average of available experimental inputs~\cite{E865:1999ker,NA482:2009pfe,NA482:2010zrc}.
For the (long-distance-dominated) transition matrix element, we can then simply write, based on the gauge and Lorentz symmetries,
\begin{equation}
\begin{split}
&\mathcal{M}_\rho\big(K^+(P)\to\pi^+(r)\gamma_\rho^*(k)\big)\\
&\equiv i\int\mathrm{d}^4x\,e^{ikx}\big\langle \pi(r)\big|\,T\big[J_\rho^\text{EM}(x)\mathcal{L}^{\Delta S=1}(0)\big]\big|K(P)\big\rangle\\
&=\frac{e}2 F(k^2)
\big[
(P-r)^2(P+r)_\rho
-(P^2-r^2)(P-r)_\rho
\big]\,;
\end{split}
\label{eq:Kpig}
\end{equation}
note that $P-r=k$ was retained for the manifestly gauge-invariant form.
The Lorentz structure of the transition is further simplified when coupled to a conserved current, such as the lepton electromagnetic current:
\begin{equation}
\mathcal{M}_\rho\big(K^+(P)\to\pi^+(r)\gamma_\rho^*(k)\big)
\stackrel{\text{eff.}}{=}e F(k^2)
k^2r_\rho\,.
\end{equation}

At tree level, the conversion $\gamma^*\to4e$ can be resolved by drawing 8 diagrams in total:
When the photon converts into an electron and positron, each of these can radiate another off-shell photon converting into another $e^+e^-$ pair; see \cref{fig:1}.
This accounts for 2 diagrams that form a gauge-invariant building block.
There are then 4 independent ways how to assign momenta to the leptons (2 for electrons and 2 for positrons).
For every such interchange, one needs to take into account a relative sign for crossing the electron lines (anticommuting two fermion fields).
The gauge-invariant building block in question is
\begin{equation}
\begin{split}
&\mathcal{M}_\gamma^\rho\big(\p,\pp;\ppp,\pppp\big)\\
    &\equiv e^3\big(I_+^{\alpha\rho}(\p,\pp;\pp)-I_-^{\rho\alpha}(\p,\pp;\p)\big)\,
    J_\alpha(\ppp,\pppp)\,,
\end{split}
\end{equation}
with
\begin{equation}
I_\pm^{\alpha\beta}(\p,\pp;p)
\equiv\bar u(\p)\gamma^\alpha\,
\frac1{\slashed P-\slashed r-\slashed p\mp m}\,\gamma^\beta v(\pp)\,,
\end{equation}
and $J_\alpha(p,q)\equiv\frac{\bar u(p)\gamma_\alpha v(q)}{(p+q)^2}$.
The lepton part of the amplitude then amounts to
\begin{align}
&\mathcal{M}^\rho_{\gamma^*\to 4e}
\equiv\mathcal{M}^\rho\big(\gamma^{*\rho}\to e^-(\p)e^+(\pp)e^-(\ppp)e^+(\pppp)\big)\notag\\
    &=\mathcal{M}_{\gamma}^\rho\big(\p,\pp;\ppp,\pppp\big)
    +\mathcal{M}_{\gamma}^\rho\big(\ppp,\pppp;\p,\pp\big)\notag\\
    &-\mathcal{M}_{\gamma}^\rho\big(\p,\pppp;\ppp,\pp\big)
    -\mathcal{M}_{\gamma}^\rho\big(\ppp,\pp;\p,\pppp\big)\,,
\end{align}
and the overall amplitude for the topology {\tz} reads
\begin{align}
    \mathcal{M}_{K\to\pi4e}^{\tz}
    &=\mathcal{M}_\rho\big(K^+(P)\to\pi^+(r)\gamma_\rho^*(k)\big)\,\frac1{k^2}\,\mathcal{M}^\rho_{\gamma^*\to 4e}\notag\\
    &=e^4F\big((P-r)^2\big)\,r_\rho\,\widetilde{\mathcal{M}}^\rho_{\gamma^*\to 4e}\,,
\end{align}
where tilde denotes the matrix element with $e=1$.

In principle, the analogous techniques employed for the calculation of the $K^+\to\pi^+\gamma^*\gamma$ within the ChPT framework can be used to obtain the form factors accompanying the Lorentz structures of the general two-photon-conversion matrix element.
In this work, only a simplified expression is utilized, based on the approximate model used originally mainly for the $K^+\to\pi^+\gamma^*\gamma$ transition.
The two-photon transition of topology {\ta} can thus be written as
\begin{align}
&\mathcal{M}_{\rho\sigma}^{(a)}\big(K(P)\to\pi(r)\gamma_\rho^*(k_1)\gamma_\sigma^*(k_2)\big)
\simeq e^2F(k_1^2)\notag\\
&\times\bigg\{
(k_1^2r_\rho-r\cdot k_1 k_{1\rho})\,\frac{(2P-k_2)_\sigma}{2P\cdot k_2-k_2^2}\notag\\
&-(k_1^2P_\rho-P\cdot k_1 k_{1\rho})\,\frac{(2r+k_2)_\sigma}{2r\cdot k_2+k_2^2}\notag\\
&+\big(k_1^2g_{\rho\sigma}-k_{1\rho}k_{1\sigma}\big)
+\kappa\big[(k_1\cdot k_2)g_{\rho\sigma}-k_{1\sigma}k_{2\rho}\big]
\bigg\}\notag\\
&+\{k_1\leftrightarrow k_2,\rho\leftrightarrow\sigma\}\,.
\label{eq:Kpgg:a}
\end{align}
This form, dependent on a single form factor (the same one as in \cref{eq:Kpig}), is particularly useful for the case when it provides the contribution to radiative corrections for the $K^+\to\pi^+\ell^+\ell^-$ decay as part of the process of measuring $F(s)$.
There, one takes one of the photons on-shell.
In the soft-photon regime, such an approximation is justified.
For the hard photons, the free parameter $|\kappa|\lesssim1$ is introduced to cover the model uncertainty.
In practice, the desired physical results do not seem to be sensitive to this parameter.
Further details can be found in Ref.~\cite{Husek:inprep}.
For our purposes, we assume that the form in \cref{eq:Kpgg:a} is good enough as an order-of-magnitude guess, at least.
Since it turns out to be numerically negligible compared to the topology (1) (with a relative suppression of one order of magnitude), we will not discuss it in greater detail here.

Finally, we look at the topology {\tb}.
The $K^+\pi^+\pi^0$ vertex can be obtained, in the framework of ChPT, from the Lagrangian~\cite{Kambor:1989tz,Cirigliano:2011ny}%
\footnote{Besides the chiral $\text{SU}(3)$ 27-plet contribution presented in \cref{eq:LG27}, there is also the octet Lagrangian $\mathcal{L}_{G_{8}p^2}^{\Delta S=1}=G_8F^4(L_\mu L^\mu)_{23}+\text{h.c.}$, which for $K^+\to\pi^+\pi^0$ leads at LO to (isospin-breaking) terms proportional to $(M_{\pi^+}^2-M_{\pi^0}^2)$ or the $\pi^0$--$\eta$ mixing angle $\epsilon^{(2)}\approx1\%$.
Such a contribution is therefore numerically small compared to the $G_{27}$ term, as $G_8/G_{27}\lesssim20$~\cite{Cirigliano:2011ny}, and it is neglected here.}
\begin{equation}
    \mathcal{L}_{G_{27}p^2}^{\Delta S=1}
    =G_{27}F^4\bigg(L_{\mu23}L_{11}^\mu+\frac23L_{\mu21}L_{13}^\mu\bigg)
    +\text{h.c.}\,,
\label{eq:LG27}
\end{equation}
with $F\approx92$\,MeV the pion decay constant, $L_{ij}^\mu=i(U^\dag \partial^\mu U)_{ij}$ and $U=\exp\frac{i\Phi}F$.
Above, the mesons are represented by the matrix $\Phi=\sum_{a=1}^8\lambda_a\phi_a$, with $\phi_a$ being the eight meson fields.
The LO matrix element for the $K^+\to\pi^+\pi^{0*}(q)$ transition following from Lagrangian \eqref{eq:LG27} is
\begin{equation}
\begin{split}
    &\mathcal{M}\big(K^+\to\pi^+\pi^{0*}(q)\big)\\
    &=-\frac i3\,F\,G_{27}\left(5M_K^2-7M_\pi^2+2q^2\right).
\end{split}
\label{eq:Kpp}
\end{equation}
On-shell and in the isospin limit, this, of course, becomes the standard result for the charged amplitude
\begin{equation}
    \mathcal{M}\big(K^+\to\pi^+\pi^{0}\big)
    =-iF\,G_{27}\,\frac53\left(M_K^2-M_\pi^2\right).
\label{eq:Kpp2}
\end{equation}
Since $B(K^+\to\pi^+\pi^0)\approx20.7\,\%$~\cite{ParticleDataGroup:2022pth}, $\frac{\Gamma_{K^+}}{M_{K^+}}\approx1.08\times10^{-16}$ and the phase space amounts to nothing else than $\frac1{8\pi}\sqrt{1-\frac{4M_\pi^2}{M_K^2}}$, we find $|G_{27}|\approx0.53\,\text{TeV}^{-2}$.
It can also be rewritten in terms of the SM parameters involved as $G_{27}=-\frac12\frac1{v^2}V_{ud}V_{us}^*\,g_{27}$, with $|V_{ud}|\approx0.974$, $|V_{us}|\approx0.224$, and $v\approx0.246$\,TeV~\cite{ParticleDataGroup:2022pth}.
Hence, we get $|g_{27}|\approx0.29$, in agreement with Refs.~\cite{Cirigliano:2003gt,Bijnens:2004ku,Cirigliano:2011ny}.
For completeness, for the virtual $\eta$ we would have
\begin{equation}
\begin{split}
    &\mathcal{M}\big(K^+\to\pi^+\eta^*(q)\big)\\
    &=-\frac i{\sqrt{3}}\,F\,G_{27}\left(3M_K^2-M_\pi^2-2q^2\right).
\end{split}
\label{eq:Kpeta}
\end{equation}

The $\pi^0\gamma\gamma$ vertex stems from the Wess--Zumino--Witten (WZW) term \cite{Witten:1983tw,Bijnens:2001bb}, which can be reduced for our application to the form
\begin{equation}
\begin{split}
\mathcal{L}_\text{WZW}^{\pi^0\gamma\gamma}
&=-\frac{N_\text{c}e^2}{24\pi^2F}\big(\pi^0+\tfrac1{\sqrt{3}}\eta\big)\epsilon^{\mu\nu\alpha\beta}(\partial_\mu A_\nu)(\partial_\alpha A_\beta)\,.
\end{split}
\label{eq:WZW}
\end{equation}
Above, we also include the $\eta$ term and one thus have (together with \cref{eq:Kpeta}) all the ingredients to include the virtual-$\eta$-exchange contribution, too.
However, this contribution can be safely neglected compared to the $\pi^0$ case and we will not consider it from now on:
It does not affect the shape of the overwhelming $\pi^0$ peak and the correction to the tail within the kinematically allowed region is rather significant but only cosmetic in the global picture when it comes to the branching-ratio estimate.
In the pion case, the Lagrangian \eqref{eq:WZW} leads to the LO transition amplitude
\begin{equation}
\mathcal{M}_{\rho\sigma}\big(\pi^{0*}\to\gamma_\rho^*(k_1)\gamma_\sigma^*(k_2)\big)
=-\frac{e^2}{4\pi^2F}\,\epsilon_{\rho\sigma\alpha\beta}k_{1}^{\alpha}k_{2}^{\beta}\,,
\label{eq:pigg}
\end{equation}
the form of which is exact for on-shell pion and photons.
For on-shell pion and off-shell photons, this can be further modulated by the doubly off-shell neutral-pion electromagnetic transition form factor $\hat{\mathcal{F}}(k_1^2,k_2^2)$, with $\hat{\mathcal{F}}(0,0)=1$.
In particular,
\begin{equation}
\begin{split}
    &\mathcal{M}_{\rho\sigma}\big(\pi^0(q)\to\gamma_\rho^*(k)\gamma_\sigma^*(q-k)\big)\\
    &\equiv i\int\mathrm{d}^4x\,e^{ikx}\big\langle0\big|\,T\big[J_\rho^\text{EM}(x)J_\sigma^\text{EM}(0)\big]\big|\pi^0(q)\big\rangle\\
    &=-\frac{e^2}{4\pi^2F}\,\hat{\mathcal{F}}\big(k^2,(q-k)^2\big)\,\epsilon_{\rho\sigma\alpha\beta}k^{\alpha}q^{\beta}\,.
\end{split}
\end{equation}
As an example, see, for instance, the model described in Ref.~\cite{Husek:2015wta}.
Further complications arise for the off-shell pion, but due to the pion pole enhancement, only the events with neutral-pion invariant mass in the vicinity of $M_{\pi^0}$ will matter.
Hence, we could safely proceed with the on-shell pion form factor and for the sake of simplicity, we will actually only consider the LO formula \eqref{eq:pigg} in what follows.
Combining \cref{eq:Kpp,eq:pigg} with a $\pi^0$-width-regulated propagator, we find the matrix element for the two-photon transition of the topology {\tb}:%
\footnote{
Let us note that instead of the LO expression \eqref{eq:Kpp} we could have used a more sophisticated form factor to be on a par with the treatment of the other two topologies.
However, enhanced by the pole, the dominant part of \cref{eq:Kpp} entering \cref{eq:Kpgg:b} is the constant \eqref{eq:Kpp2}.
And this number would just be slightly modified accounting for effects and higher-order corrections neglected here in view of considered uncertainties.
}
\begin{equation}
\begin{split}
    &\mathcal{M}_{\rho\sigma}^{(b)}(K(P)\to\pi(r)\gamma_\rho^*(k_1)\gamma_\sigma^*(k_2))\\
    &=-\frac{ie^2G_{27}}{12\pi^2}\frac{2(P-r)^2+5M_K^2-7M_\pi^2}{(P-r)^2-M_{\pi^0}^2+iM_{\pi^0}\Gamma_{\pi^0}}\,
    \epsilon_{\rho\sigma(k_1)(k_2)}\,.
\end{split}
\label{eq:Kpgg:b}
\end{equation}
Above, we used the shorthand notation $\epsilon_{\rho\sigma\alpha\beta}k_{1}^{\alpha}k_{2}^{\beta}=\epsilon_{\rho\sigma(k_1)(k_2)}$.

The two-photon-exchange topology is then the sum of the two sub-topologies from \cref{eq:Kpgg:a,eq:Kpgg:b}:
\begin{align}
\mathcal{M}_{\rho\sigma}(k_1,k_2)
&\equiv\mathcal{M}_{\rho\sigma}(K(P)\to\pi(r)\gamma_\rho^*(k_1)\gamma_\sigma^*(k_2))\notag\\
&=\mathcal{M}_{\rho\sigma}^{(a)}+\mathcal{M}_{\rho\sigma}^{(b)}\,.
\end{align}
Above, the repeating process labels were suppressed.
The two pairs of leptons are then coupled in the following way:
\begin{equation}
\begin{split}
    &\mathcal{M}_{K\to\pi4e}^{(2)}\\
    &=e^2\mathcal{M}_{K\to\pi2\gamma^*}^{\rho\sigma}(\p+\pp,\ppp+\pppp)J_\rho(\p,\pp)\,J_\sigma(\ppp,\pppp)\\
    &-e^2\mathcal{M}_{K\to\pi2\gamma^*}^{\rho\sigma}(\p+\pppp,\ppp+\pp)J_\rho(\p,\pppp)\,J_\sigma(\ppp,\pp)\,.
\end{split}
\end{equation}

Finally, we add up the one- and two-photon-exchange topologies:
\begin{equation}
    \mathcal{M}_{K\to\pi4e}
    =\mathcal{M}_{K\to\pi4e}^{\tz}+\mathcal{M}_{K\to\pi4e}^{(2)}\,.
\label{eq:1+2}
\end{equation}
As was already mentioned, it is observed that the topology {\ta} is rather suppressed compared to at least one of the other two, depending on the kinematical region, and this applies also to the (antisymmetric) interference term of topologies {\ta} and {\tz} which vanishes upon symmetric integration;%
\footnote{In other words, while the interference term is nonzero and thus should not be neglected for general application in MC, it does not contribute to the branching ratio.}
the interference of {\ta} and {\tb} is exactly zero.
This means that, in particular, we can eventually write
\begin{equation}
    \mathcal{M}_{K\to\pi4e}\simeq
        \begin{cases}
        \mathcal{M}_{K\to\pi4e}^{\tz}\,,& s\not\approx M_{\pi^0}^2\,,\\[1mm]
        \mathcal{M}_{K\to\pi4e}^{\tb}\,,& s\approx M_{\pi^0}^2\,,
        \end{cases}
\label{eq:1+2b_}
\end{equation}
with $s=(P-r)^2=(\p+\pp+\ppp+\pppp)^2$ utilized also later on, and
\begin{align}
    &\mathcal{M}_{K\to\pi4e}^{\tb}
    =-\frac{ie^4G_{27}}{12\pi^2}\frac{2s+5M_K^2-7M_\pi^2}{s-M_{\pi^0}^2+iM_{\pi^0}\Gamma_{\pi^0}}\notag\\
    &\times\big[\epsilon^{\rho\sigma(\p+\pp)(\ppp+\pppp)}J_\rho(\p,\pp)\,J_\sigma(\ppp,\pppp)\\
    &-\epsilon^{\rho\sigma(\p+\pppp)(\ppp+\pp)}J_\rho(\p,\pppp)\,J_\sigma(\ppp,\pp)\big]\,.\notag
\end{align}
Moreover, the two amplitudes in \cref{eq:1+2b_} do not interfere either, so, to ${\approx}\,10\,\%$ approximation of the branching-ratio estimate, the square of the total LO matrix element is obtained as a direct sum of the matrix elements squared of topologies {\tz} and {\tb}.
This simplifies the analytical structure significantly.
Nevertheless, for the numerical results that follow we use the complete expression given by \cref{eq:1+2}.

\section{Results}
\label{sec:results}

\begin{figure}[t]
    \centering
    \includegraphics[width=\columnwidth]{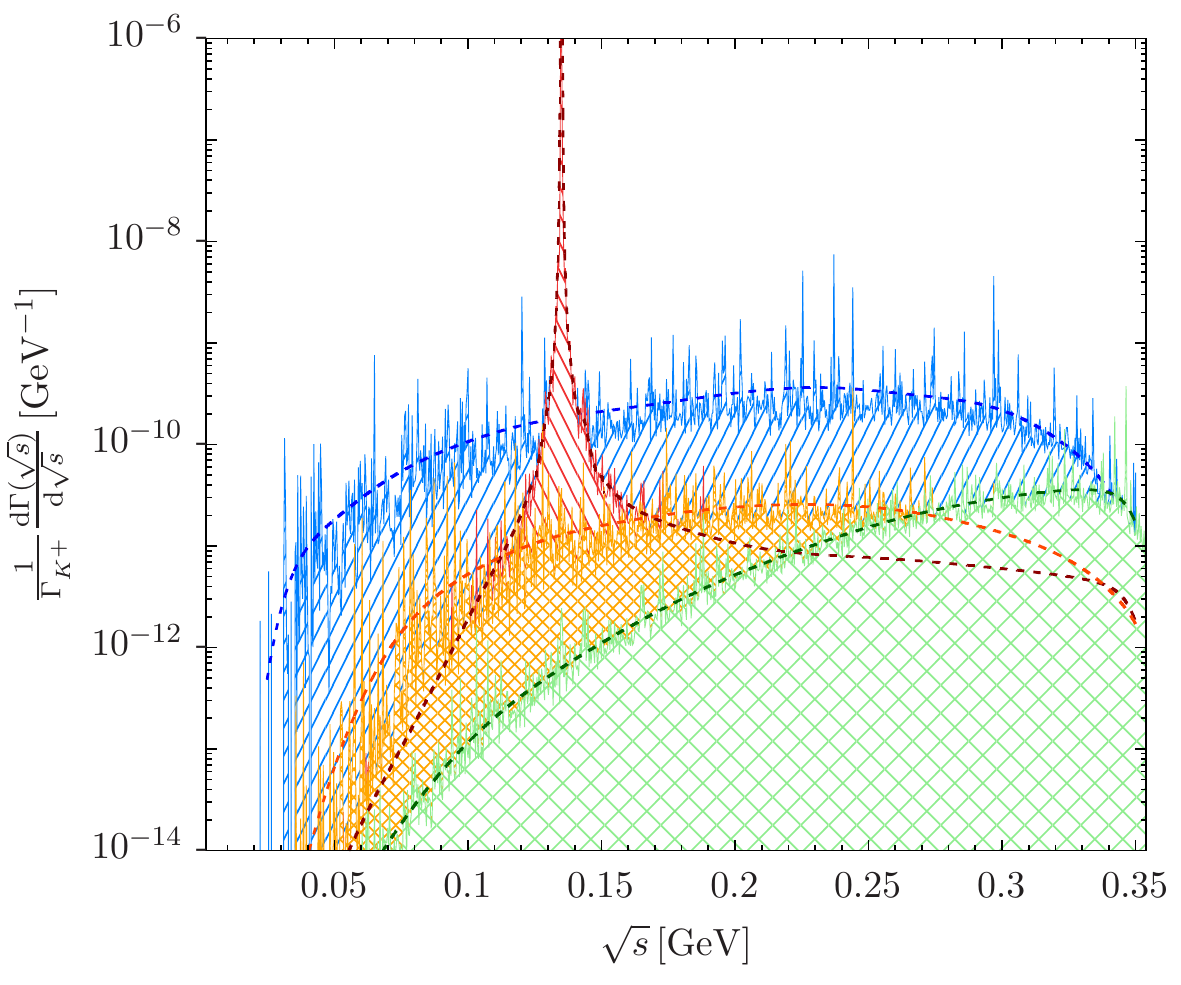}
    \caption{Monte Carlo-generated differential decay width based on the squares of the respective matrix elements constituting \cref{eq:1+2}, comparing the topologies (listed from back to front) {\tz} (in blue), {\tb} (in red), and {\ta} (in orange); in green we then show the square of the $\kappa$ term from \cref{eq:Kpgg:a} (setting $\kappa=1$) separated from {\ta} to get a handle on the uncertainty of the model leading to \cref{eq:Kpgg:a}.
    The sample used here contains $10^7$ events.
    The areas below the curves correspond directly to the respective branching ratios.
    As in \cref{eq:1+2b_}, $s=(\p+\pp+\ppp+\pppp)^2$ denotes the invariant $4e$ mass squared.}
    \label{fig:3}
\end{figure}

Having the squared matrix elements at hand, the Monte Carlo simulations can easily compare {\em relative} sizes of the respective contributions to the total branching ratio.
As expected, the topology {\tb} is by far the most dominant thanks to the pion-pole enhancement.
Its contribution to the matrix element squared and hence the overall branching ratio can be rather simply evaluated since it does not lead to interference terms with the other two topologies.
However, right outside the narrow pole region the topology {\tz} becomes prominent, so it is essential to retain it besides {\tb}; see \cref{fig:3} for the comparison of the individual contributions to the differential decay width.

Establishing the {\em absolute} branching ratio and its contributions without Monte Carlo methods requires, in the case of a 5-body decay, e.g.\ knowledge of the differential phase space in terms of at least 8 kinematical variables and subsequent multi-dimensional integration.
This can be a pretty tedious job.
At the same time, the spectra like those in \cref{fig:3} can be rather easily obtained by generating events flat-distributed in the momentum space and modulating them by the matrix element squared using, for instance, the accept/reject procedure or assigning weights given by $\overline{|\mathcal{M}|^2}$ to the events.
After binning the (accepted or weighted) events --- effectively performing the phase-space integral --- and when properly normalized, this represents the differential decay width in the chosen kinematical variable (in terms of \cref{fig:3}, this would correspond to $\frac{\mathrm{d}\Gamma(\sqrt{s})}{\mathrm{d}\sqrt{s}}$).
After the integral over this distribution is performed, one gets the branching ratio from the definition simply by normalizing the resulting width to $\Gamma_{K^+}$.
What remains is to find the proper normalization so that, for instance, \cref{fig:3} is at scale, since the information about the phase-space volume is lost during the procedure described above; see \cref{app:PS} for further details.
Inserting the (rescaled) phase-space volume $\Phi_5$ from \cref{eq:PSV} into \cref{eq:B} leads to an intuitive formula for the branching ratio:
\begin{equation}
    B=\frac1{\Gamma_{K^+}}\frac14\frac1{2M_K}\,\Phi_5\,\frac1{N}\sum_{N\,\text{events}}\overline{|\mathcal{M}|^2}\,.
\label{eq:B_}
\end{equation}
Taking the phase-space-averaged matrix element squared 
\begin{multline}
\frac1{N}\sum_{N}\overline{|\mathcal{M}_{K\to\pi4e}^{\tz}+\mathcal{M}_{K\to\pi4e}^{\ta}|^2}\\
\approx7.34(44)\times10^{-26}\,\text{MeV}^{-4}\,,    
\end{multline}
we finally arrive at
\begin{equation}
\begin{split}
    B^{{\tz}+{\ta}}(K^+\to\pi^+4e)
    &=7.2(7)\times10^{-11}\,.
\end{split}
\label{eq:BR}
\end{equation}
Taking into account the statistical uncertainty of $6\,\%$ and model uncertainty of $5\,\%$, the last number considers $10\,\%$ relative uncertainty as a conservative estimate of the final number for the branching ratio when the topology {\tb} is excluded.
This gets modified when only a limited kinematical region is considered while including all the contributions.
We find
\begin{equation}
    B(K^+\to\pi^+4e,\,\sqrt{s}>150\,\text{MeV})
    =6.0(6)\times 10^{-11}\,.
\end{equation}
A more exhaustive list of branching ratios is presented in \cref{tab:1}.
\begin{table}[t]
{\footnotesize
\setlength{\tabcolsep}{3pt}
    \renewcommand{\arraystretch}{1.5}
    \centering
    \begin{tabular}{c|c c | c}
    \toprule
         & $B(\sqrt{s}<120\,\text{MeV})$ & $B(\sqrt{s}>150\,\text{MeV})$ & $B$ \\
    \midrule
        {\tz} & $5.60\times10^{-12}$ & $5.44\times10^{-11}$ & $6.70\times10^{-11}$\\ 
        {\ta} & $3.11\times10^{-13}$ & $3.85\times10^{-12}$ & $4.60\times10^{-12}$\\ 
        {\tb} & $1.40\times10^{-13}$ & $1.97\times10^{-12}$ & $7.0(3)\times10^{-6}$ \\
    \midrule
        $\kappa$ & $7.08\times10^{-15}$ & $3.69\times10^{-12}$ & $3.72\times10^{-12}$ \\
    \midrule
        $\sum$ & $6.1(4)\times 10^{-12}$ & $6.0(6)\times 10^{-11}$ & $7.2(7)\times10^{-11}$\\
    \bottomrule
    \end{tabular}
}
    \caption{Branching ratios excluding and including the pion-pole region.
    The contributions of respective topologies are shown separately.
    They were obtained in terms of \cref{eq:B_} (restricting the sum to the given subregions) and they correspond to the relevant areas under the curves in \cref{fig:3}.
    The $\kappa$ term serves as a handle on the model uncertainty; we set $\kappa=1$.
    The total branching ratio for the topology {\tb} is taken as a product of branching ratios $B(K^+\to\pi^+\pi^0)=20.67(8)\,\%$ and $B(\pi^0\to4e)=3.38(16)\times10^{-5}$~\cite{ParticleDataGroup:2022pth}.
    The last row is a sum of the first three rows, taking the fourth row as a model uncertainty and $6\,\%$ as the statistical uncertainty.
    The rightmost value in the bottom line excludes the {\tb} topology.}
    \label{tab:1}
\end{table}

One can also check the overall consistency of the procedures used in the present work in the following way.
The total branching ratio should be
\begin{align}
    B(K^+\to\pi^+4e)
    &\simeq B(K^+\to\pi^+\pi^0)B(\pi^0\to4e)\notag\\
    &=7.0(3)\times10^{-6}\,,
\label{eq:BB}
\end{align}
since the pion-pole contribution greatly dominates; see also \cref{tab:1}.
The above expression relies on the fact that
\begin{equation}
    \int_{M-\Delta M}^{M+\Delta M}\frac{\mathrm{d}\sqrt{s}}{|s-M^2+iM\Gamma|^2}
    \simeq\frac\pi{2M}\frac1{M\Gamma}\,,\quad\Gamma\ll M\,.
\label{eq:peak}
\end{equation}
Establishing the contribution of topology {\tb} to the branching ratio using MC techniques directly is quite difficult for the physical neutral-pion width $\Gamma_{\pi^0}\approx7.8\,\text{eV}$~\cite{ParticleDataGroup:2022pth}, which makes the peak in \cref{fig:3} extremely high and narrow:
One would need a huge sample of events to start with, or possibly adapt the algorithm.
But since the integral of the peak scales as $\tfrac1\Gamma$ (see \cref{eq:peak}) and the tails are negligible, one can generate a sample with many orders of magnitude larger width, which still obeys the assumptions of \cref{eq:peak}.
In particular, if one chooses $\tilde\Gamma_{\pi^0}=10^5\Gamma_{\pi^0}$, one can rather precisely determine the ratio
\begin{equation}
    \frac{\tilde\Gamma^{\tb}}{\Gamma^{\tz}}=1.09(8)\,,
\label{eq:ratio}
\end{equation}
since now the MC event generator simulates the peak shape rather easily.
However, we already determined $B^{\tz}(K^+\to\pi^+4e)=6.70\times10^{-11}$ in \cref{tab:1} using \cref{eq:B_}.
Multiplying by the ratio \eqref{eq:ratio} and by the factor $\tilde\Gamma_{\pi^0}/\Gamma_{\pi^0}=10^5$, we find $\mbox{$B^{\tb}(K^+\to\pi^+4e)$}=7.3(7)\times10^{-6}$, as it should be based on \cref{eq:BB}.%
\footnote{
Alternatively, from the values in \cref{tab:1}, one would find
${\Gamma^{\tb}}/{\Gamma^{\tz}}=1.04(8)\times10^5$, which should be compared with \cref{eq:ratio}.
}
We thus see that the methods and values seem consistent.

\begin{acknowledgments}
I would like to thank E.~Goudzovski and A.~Shaikhiev for bringing this topic to my attention, for their valuable feedback, for the discussion regarding the MC-related results and, finally, for generating the large MC samples.
I also thank J.~Bijnens for discussions and M.~Sj\"o for proofreading.

This work is supported in part by the Swedish Research Council grants contracts no.~\mbox{2016-05996} and no.~\mbox{2019-03779}.
\end{acknowledgments}

\appendix

\section{Phase-space integral}
\label{app:PS}

The differential decay width for the $K\to\pi4e$ process is defined as
\begin{equation}
    \mathrm{d}\Gamma
    =\frac14\frac1{2M_K}\,\overline{|\mathcal{M}_{K\to\pi4e}|^2}\,\mathrm{d}\Phi_5(P;r,\p,\dots,\pppp)\,,
\label{eq:dG}
\end{equation}
where $\frac14$ is the symmetry factor accounting for 2 pairs of indistinguishable leptons in the final state and the differential phase space is
\begin{multline}
    \mathrm{d}\Phi_5(P;r,\p,\dots,\pppp)
    =(2\pi)^4\delta^{(4)}(P-r-\sum_i p_i)\\
    \times\frac{\mathrm{d}^3r}{(2\pi)^3\,2E_{r}}\,\frac{\mathrm{d}^3\p}{(2\pi)^3\,2E_{\p}}\,\dots\,\frac{\mathrm{d}^3\pppp}{(2\pi)^32\,E_{\pppp}}\,.
\end{multline}
The matrix element squared $|\mathcal{M}_{K\to\pi4e}|^2$ in \cref{eq:dG} depends, of course, on the particles' momenta, so the subsequent integral becomes largely nontrivial.
For the purposes of normalization, however, we just need to integrate the phase space on its own, taking $\overline{|\mathcal{M}_{K\to\pi4e}|^2}\to1$ in \cref{eq:dG} (and denoting such a quantity as $\mathrm{d}\hat\Gamma$).
This can be done, in the case of 5-body decay, up to 3 scalar kinematical variables.
In particular, one can write
\begin{multline}
    \frac{\mathrm{d}\Phi_5(s,s_{12},s_{34})}{\mathrm{d}s\,\mathrm{d}s_{12}\,\mathrm{d}s_{34}}
    =\phi_2(M_K^2,M_\pi^2,s)\\
    \times\frac1{(2\pi)^3}\,\phi_2(s,s_{12},s_{34})\,\phi_2(s_{12})\,\phi_2(s_{34})\,;
\label{eq:dPhisss}
\end{multline}
in general, one gets a factor $\frac1{2\pi}$ and two-body phase space accompanied by a suitable differential $\mathrm{d}s$ for every such extra branching.
Above, the (dimensionless) two-body phase space $\phi_2$ amounts to
\begin{equation}
    \phi_2(s,m_1^2,m_2^2)
    =\frac1{8\pi}\frac{\sqrt{\lambda(s,m_1^2,m_2^2)}}{s}\,,
\end{equation}
with the K\"all\'en triangle function $\lambda(a,b,c)=a^2+b^2+c^2-2ab-2ac-2bc$ and the shorthand notation $\phi_2(s)\equiv\phi_2(s,m^2,m^2)=\frac1{8\pi}\sqrt{1-4m^2/s}$.
Further, we set $s_{12}=(\p+\pp)^2$, $s_{34}=(\ppp+\pppp)^2$ and $s=(\p+\pp+\ppp+\pppp)^2$.

\begin{figure}[!t]
    \centering
    \includegraphics[width=\columnwidth]{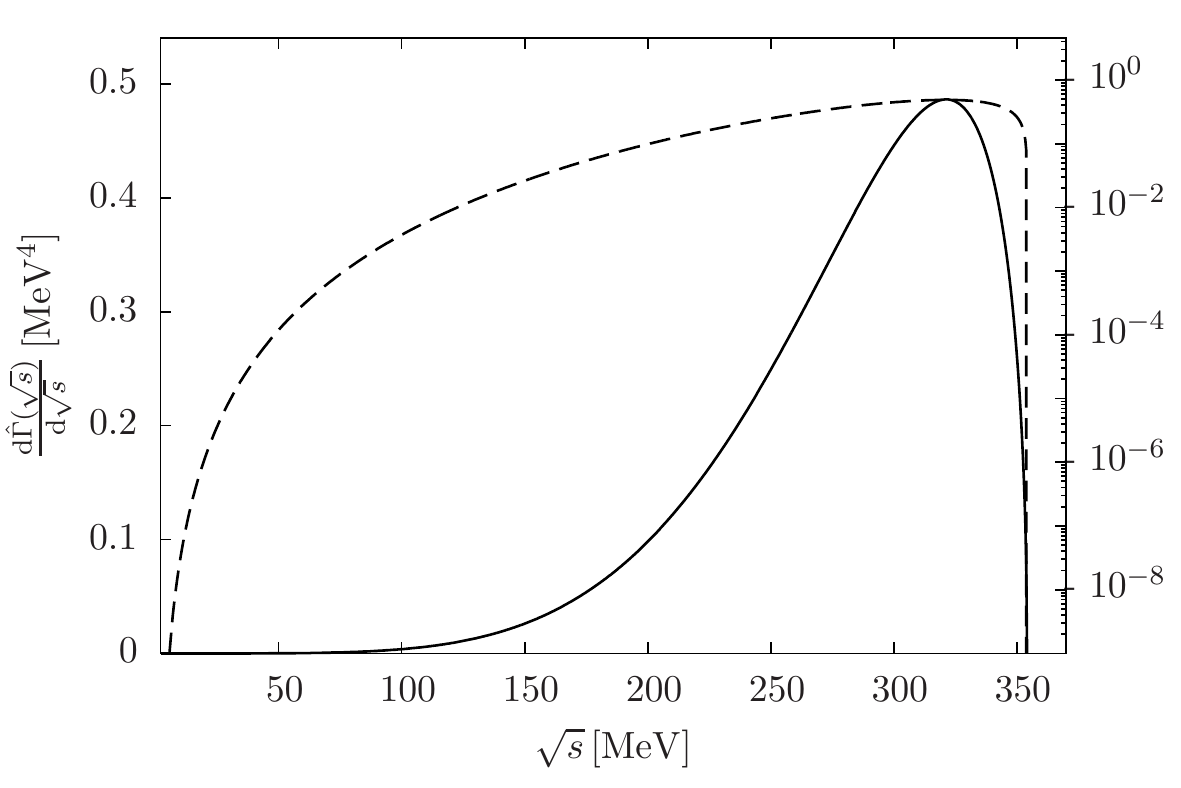}
    \caption{Differential phase space from \cref{eq:dG5}.
    It is normalized so that it corresponds to the differential decay width with matrix element squared set to unity.
    The same curve is shown both on a linear (solid) and a logarithmic (dashed) scale.
    }
    \label{fig:4}
\end{figure}

The volume of the histograms sorting the events based on specific kinematical variables is just the total number of events.
If the histogram shape corresponding to $N$ events in bins of width $w_\text{b}$ generated based on the flat phase space has a maximum $n_\text{max}\equiv n_\text{max}(N,w_\text{b})$ and is normalized so that it corresponds to (see \cref{fig:4})
\begin{equation}
\begin{split}
&\frac{\mathrm{d}\hat{\Gamma}(\sqrt{s})}{\mathrm{d}\sqrt{s}}
=\frac14\frac1{2M_K}\frac{2\sqrt{s}}{(2\pi)^3}\iint\mathrm{d}s_{12}\,\mathrm{d}s_{34}\\
&\times\phi_2(M_K^2,M_\pi^2,s)\,\phi_2(s,s_{12},s_{34})\,\phi_2(s_{12})\,\phi_2(s_{34})\,,
\end{split}
\label{eq:dG5}
\end{equation}
then $N$ events with weights according to $\overline{|\mathcal{M}_{K\to\pi4e}|^2}$ and multiplied by $\frac{\mathrm{d}\hat{\Gamma}(\sqrt{s})}{\mathrm{d}\sqrt{s}}\big|_\text{max}/n_\text{max}$ represent the desired differential decay width $\frac{\mathrm{d}{\Gamma}(\sqrt{s})}{\mathrm{d}\sqrt{s}}$; see \cref{fig:3}.
For completeness, based on \cref{eq:dG5} we find
\begin{equation}
    \frac{\mathrm{d}\hat{\Gamma}(\sqrt{s})}{\mathrm{d}\sqrt{s}}\bigg|_\text{max}
    \approx\frac{\mathrm{d}\hat{\Gamma}(321\,\text{MeV})}{\mathrm{d}\sqrt{s}}
    \approx0.486\,\text{MeV}^4\,.
\end{equation}
The integral over the described distribution then amounts to a sum over the weights $\overline{|\mathcal{M}_{K\to\pi4e}|^2}$ times an appropriate scale factor $x$ taking into account the number of generated events and the phase-space volume (while being independent of $w_\text{b}$):
\begin{equation}
    x\equiv x(N)
    =w_\text{b}\,\frac1{n_\text{max}}\frac{\mathrm{d}\hat{\Gamma}(\sqrt{s})}{\mathrm{d}\sqrt{s}}\bigg|_\text{max}\,.
\end{equation}
Alternatively, this is equivalent to normalizing to the integral of the distribution \eqref{eq:dG5},
\begin{equation}
\begin{split}
    \hat{\Gamma}
    &=\frac14\frac1{2M_K}
    \int\frac{\mathrm{d}\Phi_5(s,s_{12},s_{34})}{\mathrm{d}s\,\mathrm{d}s_{12}\,\mathrm{d}s_{34}}\,
    {\mathrm{d}s\,\mathrm{d}s_{12}\,\mathrm{d}s_{34}}\\
    &=\frac14\frac1{2M_K}\,\Phi_5
    \approx52.3\,\text{MeV}^5\,,
\end{split}
\label{eq:PSV}
\end{equation}
so the scale factor becomes simply
\begin{equation}
    x=\frac1{N}\,\hat{\Gamma}\,.
\end{equation}
Hence, in general, the branching ratio $B$ can be obtained as
\begin{equation}
    B=\frac{\hat{\Gamma}}{\Gamma_0}\frac1{N}\sum_{N\,\text{events}}\overline{|\mathcal{M}|^2}\,,
\label{eq:B}
\end{equation}
which represents a rescaled average of the matrix element squared over the phase space (with events randomly and evenly distributed in the momentum space) and is used to get the values in \cref{tab:1}.


\let\raggedright

\providecommand{\href}[2]{#2}\begingroup\raggedright\endgroup

\end{document}